\documentclass[10pt,aps,prl,twocolumn,preprintnumbers,nofootinbib,amsmath,amssymb,superscriptaddress]{revtex4-2}
\usepackage{orcidlink}

\usepackage{graphicx}
\usepackage{dcolumn}
\usepackage{bm}
\usepackage{color}
\usepackage{xcolor}

\hypersetup{
  colorlinks=true,
  linkcolor=blue,
  citecolor=red,
  urlcolor=magenta,
  breaklinks=true
}

\begin{document}
\date{\today}


\title{Role of inertia on the performance of Brownian gyrators}

\author{Thalyta T. Martins\orcidlink{https://orcid.org/0000-0003-2113-5468}}%
\author{Ines Ben-Yedder}%
\author{Alex Fontana \orcidlink{https://orcid.org/0000-0001-5522-3584}}%
\author{Loïc Rondin \orcidlink{https://orcid.org/0000-0002-4833-2886}}%
\email{loic.rondin@universite-paris-saclay.fr}
\affiliation{Université Paris-Saclay, ENS Paris-Saclay, CNRS, CentraleSupelec, LuMIn, 91400, Orsay, France.}

\date{\today}

\begin{abstract}
{Understanding the role of inertia in nanoscale heat transport is fundamental to the design of efficient nano-thermodynamics systems.
In this work, we experimentally address the non-equilibrium dynamics of a Brownian gyrator, a paradigmatic model for nano-heat machines, that converts heat flow between two thermal baths into steady-state rotation. Using an optically levitated nanoparticle in a controlled vacuum environment, we study the transition from overdamped to underdamped dynamics of the gyrator. We demonstrate that, while the spatial signature of the non-equilibrium steady state vanishes as damping decreases, the rotational dynamics and energetics are optimized at a critical damping. 
Our findings reveal the importance of inertia for maximising the performance of nanoscale machines and provide fundamental insights into the design of efficient nano heat engines and processes.}

\end{abstract}

\maketitle
Understanding and enhancing the conversion between heat and mechanical energy is a cornerstone of thermodynamics. This principle has led to the development of heat engines, which have fundamentally transformed society.
In recent decades, attention has turned to extending these ideas to the nano- and microscale, where thermal fluctuations play a dominant role. In this context, the nano-heat engine has emerged as a paradigmatic system and has recently been experimentally realized~\cite{blickle2012realization, martinez2016brownian, rossnagel2016Singleatom}, opening new avenues for both fundamental thermodynamics and the design of practical, small-scale engines.
Of particular importance are recent efforts to explore such systems in the underdamped regime~\cite{dechant2015AllOptical, li2024realization, message2025Extremetemperature}, where inertia cannot be neglected. This regime is crucial, since it represents the most general case in nanothermodynamics. Even for heavily damped systems, the full dynamics must often be accounted for, specifically while interested in energetic properties~\cite{arold2018Heat}. Moreover, it offers a unique opportunity to investigate nano-heat engines in the quantum regime~\cite{dechant2015AllOptical}, leveraging the development of nano-mechanical systems in the quantum regime~\cite{delic2020Cooling,tebbenjohanns2021Quantum,magrini2021Realtime,kamba2023Revealing}.

Due to its simplicity and the richness of its physical properties, the Brownian gyrator constitutes a particularly well-suited model for studying nanoscale thermal machines. This minimal system captures the essential features of a non-equilibrium steady state (NESS) in which a particle coupled to two orthogonal heat baths exhibits autonomous steady rotation~\cite{filliger2007brownian}. The simplicity of this model has triggered a broad body of literature, including experimental realizations in the overdamped regime using colloidal particles in optical tweezers~\cite{argun2017experimental, lin2022stochastic}, as well as electronic circuits~\cite{chiang2017electrical}.
However, the role of inertia—and thus the underdamped regime—has so far been experimentally overlooked.

In the present work, we use an optically levitated particle in rarefied air to explore the dynamics of the Brownian gyrator and assess the impact of inertia on its rotational dynamics and the associated energetic properties.  
\begin{figure}[ht]
    \centering
    \includegraphics[width=.6\textwidth]{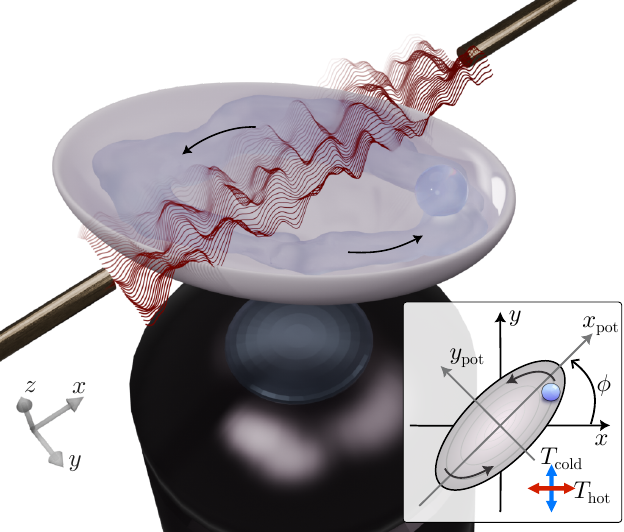}
    \caption{\textbf{Brownian gyrator setup.} A laser focused through an objective traps a glass particle in an asymmetric harmonic potential. A pair of electrodes generates a directional effective temperature $T_{\mathrm{hot}}$ along the $x$ axis (red waves) larger than the temperature along the $y-$axis $T_\text{cold}$. The potential principal axes $\bm{x}_\mathrm{pot}$ and $\bm{y}_\mathrm{pot}$ are tilted by an angle $\phi=\pi/4$ relative to the lab frame $(\bm{x}, \bm{y})$ (inset).  The broken system symmetry results in a circular counterclockwise motion represented by the light blue flow and the black arrows.}
    \label{fig:scheme}
\end{figure}

\begin{figure*}[htbp!]
    \centering
    \includegraphics[width=.95 \textwidth]{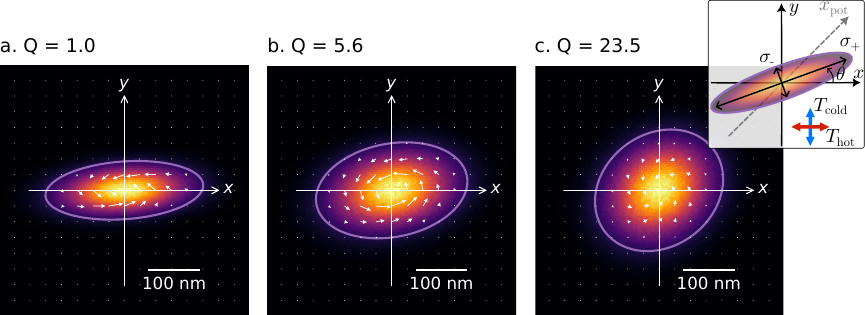}
    \caption{\textbf{Gyrator's PDF.} Position's PDF and probability current (white arrows) in the $(x, y)$ plane, highlighting the counterclockwise rotation of the particle, for quality-factor a. $\mathsf Q = 1$ (overdamped, $p_\text{gas}=220 \ \mathrm{mbar}$);  b. $\mathsf Q = 5.6 $ (critical damping, $p_\text{gas}= 33\ \mathrm{mbar}$) and c. $\mathsf Q = 23.5$ (underdamped regime, $p_\text{gas}=8 \ \mathrm{mbar}$). The PDF is fitted with a 2D Gaussian function; highlighted by the 4$\sigma$ iso-contour (purple ellipse). We extract from the fit the PDF tilt angle $\theta$ and its major and minor axes standard deviations  $\sigma_+$ and $\sigma_-$ (inset).
    }
    \label{fig:pdf} 
\end{figure*}

The working principle of the Brownian gyrator is depicted in the inset of Figure~\ref{fig:scheme}. It consists of a Brownian particle trapped in a two-dimensional asymmetric harmonic potential coupled to two orthogonal thermal reservoirs. The hotter bath, at temperature $T_\text{hot}$, is aligned with the $\bm{x}$-direction, while the colder bath (temperature $T_\text{cold}$) is aligned with the $\bm{y}$-direction. The principal axes of the harmonic potential, $\bm{x}_\text{pot}$ and $\bm{y}_\text{pot}$  are tilted by an angle  $\phi$ with respect to the lab frame $(\bm{x}, \bm{y})$. 
The harmonic trap is thus characterized by trap stifnesses $k_{x_\text{pot}}= k - u$  and  $k_{y_\text{pot}}= k + u$, where we introduce $k$ as the average trap stiffness and $u\geq 0$ the potential asymmetry. 
The particle's dynamics, in the lab-frame, are then described by the Langevin equations:
\begin{equation}
\ddot{\bm {r}}(t)
+ \Gamma \dot{\bm{r}}(t)
+ \frac{1}{m}\mathbf{K}\bm {r}(t)
= \frac{1}{m}\bm{\xi}(t),
\end{equation}
where 
$$
\bm{r}(t)=
\begin{pmatrix}
x(t)\\
y(t)
\end{pmatrix},
\qquad
\mathbf{K}=
\begin{pmatrix}
k - u \cos 2\phi & -u\sin 2\phi\\
-u \sin 2\phi & k + u  \cos 2\phi 
\end{pmatrix},
$$
$\boldsymbol{\xi}$ is the thermal Gaussian white noise with 
$$
\langle \bm{\xi}(t)\bm{\xi}^T(t') \rangle
= 2 m \Gamma k_B
\begin{pmatrix}
  T_\text{hot} & 0\\
  0 & T_\text{cold}
\end{pmatrix}
\delta(t-t')\, , 
$$
$m$ is the particle mass and $\Gamma$ is the system damping.
Note that both the potential symmetry ($u\neq0)$ and misalignment between the potential and heat-baths ($\phi\neq 0$) are essential for the system  
to break symmetry and to enter a NESS characterized by rotational probability currents.
The emergence of rotation can be qualitatively understood by decomposing the motion along the anisotropic temperature axes. 
Considering a Brownian particle on the $y$-axis (cold axis), the restoring force associated with the potential favors a clockwise rotation  (see inset in Figure~\ref{fig:scheme}). Conversely, for the particle on the $x$-axis (hot axis), a counterclockwise motion is expected. Since the particle diffuses in the high-temperature direction more extensively, it should, at the end, have a net counterclockwise circulation. Inverting hot and cold baths ($\phi \rightarrow -\phi$) reverts the effect, leading to a net clockwise circulation~\cite{argun2017experimental}.

To experimentally address the impact of inertia on the Brownian gyrator, we use an optically levitated particle. This system constitutes a perfect testbed for the study of in and out of equilibrium nanothermodynamics~\cite{gieseler2018Levitated,gonzalez-ballestero2021Levitodynamics}, by providing unique control on the trapping potentials, heat baths, and on the damping coefficient $\Gamma$~\cite{rondin2017Direct,raynal2023Shortcuts}.
The experimental apparatus is depicted in Figure~\ref{fig:scheme}.  We optically trap a 106~nm radius silica particle inside a vacuum chamber using a tightly focused infrared laser ($\lambda_t = 1550 \ \mathrm{nm}$). The full particle dynamics is recorded by an interferometric detection of the forward-scattered trapping light using a quadrant photodiode. The detection is related to the lab-frame and defines the $(\bm x,\bm y)$ frame. 
The trap asymmetry in the $(x_\text{pot}, y_\text{pot})$ plane is induced by the laser light linear polarization, which slightly increases the beam waist along the polarization direction, reducing the trap stiffness along this axis~\cite{gieseler2012subkelvin}. Experimentally, we measured an averaged trap stiffness $k=4.81$~pN/µm and an asymmetry $u=0.8$~pN/µm.
We fix the potential orientation to $\phi=\pi/4$ using a polarization half‑wave plate, to maximize the circulation effect (see Supplementary Information).  
To generate the anisotropic bath temperature, we apply a directional noisy electric force to the electrically charged particle~\cite{martinez2013Effective,militaru2021kovacs} thanks to a pair of electrodes. These are aligned along the $x$-axis (see Figure ~\ref{fig:scheme}).
Consequently, the effective ${x}$‑axis temperature $T_\mathrm{hot}$ is increased above the ${y}$‑axis temperature $T_\mathrm{cold}$ which is kept at room temperature $T_\text{cold} = 295$~K.
In the following, the ${x}$‑axis temperature is calibrated to  $T_\mathrm{hot} = 2360 \pm 40$~K for the whole experiment.
Details of these calibration procedures are provided in the~\textit{Supplementary Information}. 
To characterize the role of inertia on this Brownian gyrator, we vary the system damping $\Gamma$ by tuning the gas pressure inside the vacuum chamber~\cite{rondin2017Direct}. The transition from the overdamped regime, where inertia can be neglected, to the underdamped regime is of special interest. We quantify the damping regime by introducing the mechanical quality factor $\mathsf Q=\Omega_0/\Gamma$ of the levitated particle, where $\Omega_0=\sqrt{k/m}$ is the average natural frequency of the trap. 
Thus, tuning the chamber pressures from 220~mbar to $\approx 8$~mbar allows control of the system from $\mathsf Q\approx 1$ (overdamped) to $\mathsf Q\approx 30$ (underdamped).

\begin{figure}[htpb!]
    \centering
    \includegraphics[width=.45\textwidth]{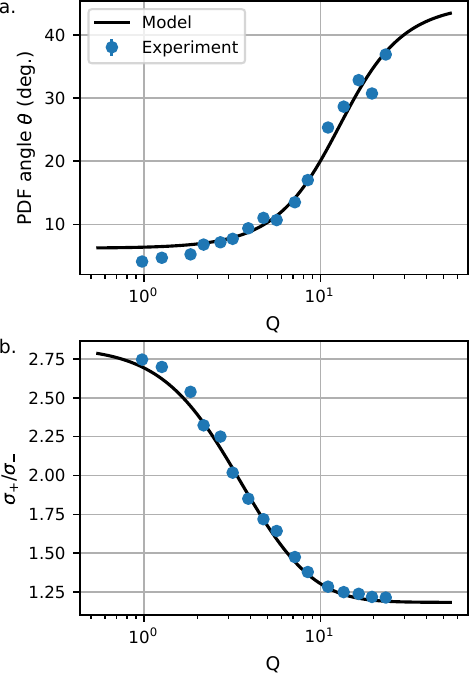}
    \caption{\textbf{Properties of the gyrator's PDF.} a. Angle ($\theta$) and b. aspect ratio of the standard deviations ($\sigma_+/\sigma_-$) along the principal axes of the PDF for different damping values. Experimental data (blue dots) computed from three realizations of the gyrator, and theoretical predictions (black curves) for the experimental parameter set. The error bars are smaller than the dot size.}
    \label{fig:PDF_tilt} 
\end{figure}

To analyze the system’s non-equilibrium behavior, we examine the steady-state probability density function (PDF) of the particle's position, $p(x,y)$, obtained by normalizing the two-dimensional histogram of the recorded trajectory $(x(t),y(t))$, as shown in Figure ~\ref{fig:pdf}. 
A signature of the Brownian gyrator appears as an elongation and tilt of the PDF toward the direction of the hot bath~\cite{argun2017experimental}. We clearly observe this effect for overdamped dynamics ($\mathsf Q \approx 1$, Figure ~\ref{fig:pdf}a), demonstrating the Brownian gyrator for our levitated system. However, as the damping decreases and the system becomes increasingly underdamped ($\mathsf Q \gg 1$), the PDF tilt and elongation gradually fade. 
Ultimately, as shown in Figure ~\ref{fig:pdf}c, for $\mathsf Q = 23.5$, the PDF aligns with the potential, as would be the case for an equilibrium state. 
To quantitatively characterize these effects, we fit the shape of the PDF with a 2D Gaussian function as highlighted by the purple ellipses in Figure ~\ref{fig:pdf}. From these fits, we extract the PDF tilt angle $\theta$, as well as its standard deviations along its principal axes, $\sigma_{+}$ and $\sigma_{-}$ (see inset of Figure ~\ref{fig:pdf}). 
Figure~\ref{fig:PDF_tilt} shows the angle $\theta$ and the aspect ratio of the standard deviations $\sigma_{+}/\sigma_{-}$ (b), characterizing its elongation. These results thus confirm that the Brownian-gyrator signature in the PDF vanishes in the underdamped regime. Interestingly, in the vanishing damping regime, the nonequilibrium steady state resembles a thermal equilibrium with an averaged temperature $T_\text{eff}=(T_\mathrm{hot}+T_\mathrm{cold})/2$. Indeed, the standard deviation ratio converges to
    \begin{equation}
      \left(\dfrac{\sigma_+}{\sigma_-}\right) \rightarrow \sqrt{\frac{k_{x_\text{pot}}}{k_{y_\text{pot}}}}\, , 
      \label{eq:elong_UD}
    \end{equation}
consistent with the equipartition theorem, and the angle of the major axis approaches the original angle of the potential $\theta \rightarrow \phi = \pi/4$. This effect is attributed to the fact that in the underdamped regime, the system's dynamics are governed by the slow evolution of the kinetic energy rather than spatial diffusion, leading the total system energy to relax to the averaged thermal energy.
Moreover, our experimental findings are strengthened by the excellent agreement with theoretical predictions, shown as black curves in Figure~\ref{fig:PDF_tilt}. Specifically, we compute the expected theoretical tilt angle $\theta$ and elongation ratio $\sigma_{+}/\sigma_{-}$ of the PDF by diagonalizing the covariance matrix of the steady-state marginal distribution in the $(x,y)$ space~\cite{mancois2018Twotemperature,bae2021inertial} (see Methods). 

Thus, the analysis of the PDF indicates that inertia strongly affects the dynamics of the Brownian gyrator, masking its non-equilibrium nature in the underdamped limit and making this thermal machine model not well-suited for exploring nano-heat engines in this regime. 
Nevertheless, beyond the PDF structure, an essential aspect of the Brownian gyrator is the persistent rotational component in its dynamics.
In this context, a natural quantity to compute is the particle rotation probability current~\cite{argun2017experimental,lin2022stochastic,seifert2025Stochastic}:
\begin{equation}
    \bm{j}(x', y') = \langle \bm{v} \rangle|_{(x',y')} \, p(x', y')\, ,
\end{equation}
where $\langle \bm{v} \rangle|_{(x',y')}$ 
is the local mean velocity, computed by final differences of the particle dynamics $({x}(t), {y}(t))$, for the particle at the position $(x',y')$~\cite{argun2017experimental}. This probability current is represented by the white arrows in Figure ~\ref{fig:pdf}. It clearly reveals a counterclockwise circulation, as expected for the Brownian gyrator. However, as shown in Figure ~\ref{fig:pdf}c, the current amplitude decreases at large  $\mathsf Q$, highlighting once again the inefficiency of the Brownian gyrator for highly underdamped systems.

\begin{figure}[htpb]
    \centering
    \includegraphics[width=.45\textwidth]{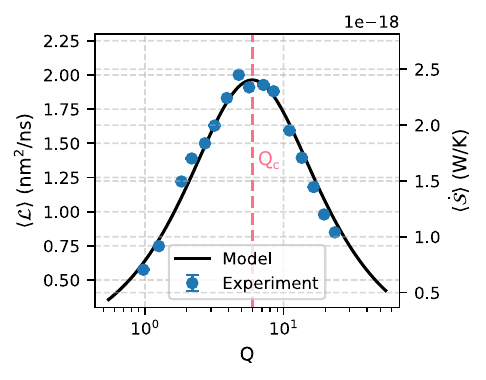}
    \caption{\textbf{Angular momentum and entropy creation}.  Experimental data (blue dots), as a function of resonator quality factor, computed from three realizations of the gyrator, and theoretical predictions (black curves) for the experimental parameter set. The error bars are smaller than the dot size. The pink dashed line represents the critical damping $\mathsf Q_c=\dfrac{k}{u}$.}
    \label{fig:momentum} 
\end{figure}
To properly characterize the rotational efficiency of the Brownian gyrator, we introduce the gyrator angular momentum defined as $\mathcal{\bm L} = \bm r \times \bm v$. 
Interestingly, this quantity links directly to both the heat absorbed from the hot bath and the total entropy production, defined as the sum of the system's and medium's entropy, through (see Methods)~\cite{bae2021inertial}:
\begin{equation}
  \langle \dot{Q}_h \rangle =  \frac{u}{2} \langle \mathcal L \rangle, \quad \mathrm{and} \quad \langle \dot{\mathcal{S}} \rangle = -\left( \frac{1}{T_\mathrm{hot}}- \frac{1}{T_\mathrm{cold}}\right) \frac{u}{2} \langle \mathcal L \rangle.
\label{eq:Q_h}
\end{equation}
We compute $\langle \mathcal L \rangle$ directly from the experimentally measured particle's positions $\bm r(t)$ and velocity $\bm v(t)$ numerically computed by time difference of $\bm r$. 
The resulting angular momentum and corresponding entropy production as a function of damping are shown in Figure~\ref{fig:momentum}. These quantities display a maximum at a specific damping, revealing an optimal damping regime for efficient heat exchange between the particle and the heat baths. At high damping ($\mathsf Q \leq 1$), the particle is heavily damped by the environment, quenching its rotation.
Conversely, in the deep underdamped regime ($\mathsf Q\gg 1$), the particle’s coherent oscillations lead to an averaging of the two thermal baths, effectively relaxing the NESS toward a mean-temperature equilibrium-like state. A peak is then observed at a critical damping $\mathsf Q_c\approx 6$ (dashed line in Figure~\ref{fig:momentum}).
Using the covariance matrix of the system (see Methods), we can compute the theoretical average angular momentum:
\begin{equation}
    \langle \mathcal L \rangle  = u \Gamma \frac{ (T_{\mathrm{hot}} - T_{\mathrm{cold}})}{m k \Gamma^2 + u^2}\, .
    \label{eq:L_theo}
\end{equation}
The results provide once again an excellent agreement with experimental data (Fig.~\ref{fig:momentum}) and show that the critical damping writes $\mathsf Q_c=\dfrac{k}{u}$, once again agreeing with experimental findings, and highlighting the importance of the potential asymmetry $u$ on the observed maxima. 
Finally, these results highlight that to maximize heat transfer or entropy production, the system’s friction must be tuned to an optimal value. This effect is also expected for a Brownian gyrator acting against an external torque, which can then be used as a practical heat engine producing work~\cite{lin2022stochastic}. We also expect that, in that case, tuning the damping will help optimize the heat engine's efficiency~\cite {bae2021inertial}.


In conclusion, we have experimentally implemented a Brownian gyrator and characterized its rotational efficiency across the transition from the overdamped to the underdamped regime, using a levitated nanoparticle. 
Our results demonstrate a fundamental shift in the manifestation of the non-equilibrium steady states: as inertia becomes dominant, the spatial asymmetry of the probability density function, a hallmark of the overdamped gyrator, effectively vanishes. 
Furthermore, the energetic properties of the gyrator are maximized at an intermediate damping. The observation of a maximum in angular momentum and entropy production indicates that damping plays a critical role in the efficiency of heat transfer between the thermal reservoirs and nanoscale engines.
This finding thus establishes the importance of inertia in mesoscopic energy transport and offers an interesting framework and experimental platform for optimizing thermodynamic processes at the nanoscale~\cite{patron2024Minimuma,baldassarri2020Engineered}.

\newpage

\section*{Methods}
\subsection*{Probability distribution function}
The theory of the Brownian gyrator in the general damping case has been discussed in references~\cite{mancois2018Twotemperature,bae2021inertial}. 

Following the definition of the main text, the dynamics of the particle can be cast under the  Langevin equation 
\begin{equation}
\ddot{\bm {r}}(t)
+ \Gamma \dot{\bm{r}}(t)
+ \frac{1}{m}{\mathbf{K}}\bm {r}(t)
= \frac{1}{m}\bm{\xi}(t),
\end{equation}
where 
$$
\bm{r}(t)=
\begin{pmatrix}
x(t)\\
y(t)
\end{pmatrix},
\qquad
\mathbf{K}=
\begin{pmatrix}
k - u \cos 2\phi & -u\sin 2\phi\\
-u \sin 2\phi & k + u  \cos 2\phi 
\end{pmatrix},
$$
$\bm{\xi}$ is the thermal Gaussian white noise with 
$$
\langle \boldsymbol{\xi}(t)\boldsymbol{\xi}^T(t') \rangle
= 2 m \Gamma k_B
\begin{pmatrix}
  T_\text{hot} & 0\\
  0 & T_\text{cold}
\end{pmatrix}
\delta(t-t')\, , 
$$
$m$ is the particle mass and $\Gamma$ is the system damping, $\phi$ the orientation of the potential, $k$ the average stiffness, and $u$ the potential asymmetry. 

Given the matrices $\mathbf K$ and $\bm{\xi}$, one can compute an analytical form of the covariance matrix $\mathbf{C_\phi}$ at steady state~\cite{bae2021inertial,mancois2018Twotemperature}. For instance, for $\phi=\pi/4$, we found 
\begin{widetext}
    \begin{equation}
        \mathbf{C_\frac{\pi}{4}} = \begin{pmatrix}
        \frac{2 k \left(\Gamma^{2} k m T_\text{hot}  + \bar T u^{2}\right) - \Gamma^{2} m u^{2} \Delta T}{2\left(k^{2} - u^{2}\right) \left(\Gamma^{2} k m + u^{2}\right)} 
        & \frac{u \bar T}{ \left(u^{2} - k^{2} \right)} & 0 & \frac{\Gamma u \Delta T}{2 \left(\Gamma^{2} k m + u^{2}\right)}\\
         \frac{u \bar T}{u^{2}- k^{2}} & 
         \frac{2 k \left(\Gamma^{2} k m T_\text{cold}  + \bar T u^{2}\right) + \Gamma^{2} m u^{2} \Delta T}{2\left(k^{2} - u^{2}\right) \left(\Gamma^{2} k m + u^{2}\right)} 
         & -\frac{\Gamma u \Delta T}{2 \left(\Gamma^{2} k m + u^{2}\right)} & 0\\
        0 & 
        -\frac{\Gamma u \Delta T}{2 \left(\Gamma^{2} k m + u^{2}\right)}  
       & \frac{\Gamma^{2} T_\text{hot} k m + \bar T u^{2}}{ m \left(\Gamma^{2} k m + u^{2}\right)} & 0\\
        \frac{\Gamma u \Delta T}{2 \left(\Gamma^{2} k m + u^{2}\right)} & 0 & 0& 
        \frac{\Gamma^{2} T_\text{cold} k m + \bar T u^{2}}{ m \left(\Gamma^{2} k m + u^{2}\right)}\end{pmatrix}
        \label{eq:matC}
    \end{equation}
\end{widetext}
where $\bar T= k_B \dfrac{T_\text{hot}+T_\text{cold}}{2}$ and $\Delta T=k_B\left(T_\text{hot}-T_\text{cold}\right)$.

In the steady state, the phase-space probability distribution function (PDF) $ p(\bm r, \bm v, t) $ is Gaussian and can be written in the form:
\begin{equation}
    p_z(\boldsymbol{z}) = \frac{1}{(2\pi)^2 \sqrt{\det \boldsymbol{C}}} \exp\left( -\frac{1}{2} \boldsymbol{z}^\mathrm{T} \boldsymbol{C}^{-1} \boldsymbol{z} \right),
    \label{eq:PDF}
\end{equation}
where $ \boldsymbol{z} = \begin{pmatrix} x & y & v_x & v_y \end{pmatrix}^T $ is the state vector in phase space.

In the main text, we analyze the tilt $\theta$ and the standard deviations along the main axes, $\sigma_\pm$, of the position PDF $p(x,y)$ 
\begin{equation}
    p(\bm r ) = \frac{1}{(2\pi)^2 \sqrt{\det \boldsymbol{C_{r}}}} \exp\left( -\frac{1}{2} \bm{r}^\mathrm{T} \boldsymbol{C_r}^{-1} \bm{r} \right),
    \label{eq:PDF_r}
\end{equation}
where $\mathbf{C_{r}}$ is the marginal covariance matrix to the $(x,y)$ phase space. 

We finally extract the tilt and the standard deviation along the main axes of this 2D Gaussian marginal PDF by diagonalizing the covariance matrix: the eigenvectors determine the orientation $\theta$ of the major and minor axes, while the eigenvalues give their respective standard deviation $\sigma_\pm$. The results are used as a theory curve in Figure~3 of the main text. 

\subsection*{Angular momentum, heat exchange and entropy production}

In the main text, we define the angular momentum as 
\begin{equation}
    \bm{\mathcal L} (\bm{r}, \bm{v}, t) = \bm{r}(t) \times \bm{v}(t) \, . 
    \label{eq:ang_momentum}
\end{equation} 
Using the definition of the covariance matrix (eq.~\ref{eq:matC}), we find the average steady-state value of the angular-momentum
\begin{equation}
    \langle \mathcal L \rangle  =  \langle xv_y-yv_x \rangle = u \Gamma \frac{ (T_{\mathrm{hot}} - T_{\mathrm{cold}})}{m k \Gamma^2 + u^2},
    \label{eq:L_theo_m}
\end{equation}

In the steady state, the average heat rates $\langle \dot{Q} \rangle$ exchanged with the hot and cold baths are~\cite{bae2021inertial}:
\begin{equation}
    \langle \dot{Q}_\text{hot} \rangle = - \langle \dot{Q}_\text{cold} \rangle = \frac{u}{2} \langle \mathcal{L} \rangle.
\end{equation}
The resulting steady-state total entropy production rate is:
\begin{equation}
    \langle \dot{\mathcal{S}} \rangle = -\frac{\langle \dot{Q}_\text{hot} \rangle}{T_\text{hot}} - \frac{\langle \dot{Q}_\text{cold} \rangle}{T_\text{cold}} = -\frac{u }{2} \left( \frac{1}{T_\text{hot}} - \frac{1}{T_\text{cold}} \right) \langle \mathcal{L} \rangle.
\end{equation}

\section*{Supplementary Information}
The article is accompanied by a supplementary information file that discusses experimental details, calibration procedures, and the impact of the potential orientation on the Brownian Gyrator. 

\section*{Acknowledgment}
This work is supported by the Paris Île-de-France Region within the framework of the DIM SIRTEQ, and by the ANR FENNEC project (ANR-23-CE30-0042). We thank Carlos A. Plata and Antonio Prados for their insightful discussions and comments on the manuscript.

\bibliography{refs}

@article{militaru2021kovacs,
  title={Kovacs memory effect with an optically levitated nanoparticle},
  author={Militaru, Andrei and Lasanta, Antonio and Frimmer, Martin and Bonilla, Luis L and Novotny, Lukas and Rica, Ra{\'u}l A},
  journal={Physical Review Letters},
  volume={127},
  number={13},
  pages={130603},
  year={2021},
  publisher={APS}
}

@article{bae2021inertial,
  title={Inertial effects on the Brownian gyrator},
  author={Bae, Youngkyoung and Lee, Sangyun and Kim, Juin and Jeong, Hawoong},
  journal={Physical Review E},
  volume={103},
  number={3},
  pages={032148},
  year={2021},
  publisher={APS},
  url={https://doi.org/10.1103/PhysRevE.103.032148}
}

@article{filliger2007brownian,
  title={Brownian gyrator: A minimal heat engine on the nanoscale},
  author={Filliger, Roger and Reimann, Peter},
  journal={Physical review letters},
  volume={99},
  number={23},
  pages={230602},
  year={2007},
  publisher={APS},
  url={https://doi.org/10.1103/PhysRevLett.99.230602}
}

@article{welch1967use,
  title={The use of fast Fourier transform for the estimation of power spectra: a method based on time averaging over short, modified periodograms},
  author={Welch, Peter},
  journal={IEEE Transactions on audio and electroacoustics},
  volume={15},
  number={2},
  pages={70--73},
  year={1967},
  publisher={IEEE},
  url = {https://doi.org/10.1109/TAU.1967.1161901}
}

@article{mancois2018Twotemperature,
  title = {Two-Temperature {{Brownian}} Dynamics of a Particle in a Confining
           Potential},
  author = {Mancois, Vincent and Marcos, Bruno and Viot, Pascal and Wilkowski,
            David},
  year = 2018,
  month = may,
  journal = {Phys. Rev. E},
  volume = {97},
  number = {5},
  pages = {052121},
  publisher = {American Physical Society},
  doi = {10.1103/PhysRevE.97.052121},
  urldate = {2024-10-23},
}

@article{lin2022stochastic,
  title = {Stochastic currents and efficiency in an autonomous heat engine},
  author = {Lin, Wenqi and Liao, Yi-Hung and Lai, Pik-Yin and Jun, Yonggun},
  journal = {Physical Review E},
  volume = {106},
  number = {2},
  pages = {L022106},
  year = {2022},
  publisher = {APS},
}

@article{gieseler2012subkelvin,
  title = {Subkelvin parametric feedback cooling of a laser-trapped nanoparticle
           },
  author = {Gieseler, Jan and Deutsch, Bradley and Quidant, Romain and Novotny,
            Lukas},
  journal = {Physical review letters},
  volume = {109},
  number = {10},
  pages = {103603},
  year = {2012},
  publisher = {APS},
}

@article{argun2017experimental,
  title = {Experimental realization of a minimal microscopic heat engine},
  author = {Argun, Aykut and Soni, Jalpa and Dabelow, Lennart and Bo, Stefano
            and Pesce, Giuseppe and Eichhorn, Ralf and Volpe, Giovanni},
  journal = {Physical Review E},
  volume = {96},
  number = {5},
  pages = {052106},
  year = {2017},
  publisher = {APS},
  url = {https://doi.org/10.1103/PhysRevE.96.052106},
}

@article{blickle2012realization,
  title = {Realization of a micrometre-sized stochastic heat engine},
  author = {Blickle, Valentin and Bechinger, Clemens},
  journal = {Nature Physics},
  volume = {8},
  number = {2},
  pages = {143--146},
  year = {2012},
  publisher = {Nature Publishing Group UK London},
}

@article{martinez2016brownian,
  title = {Brownian carnot engine},
  author = {Mart{\'\i}nez, Ignacio A and Rold{\'a}n, {\'E}dgar and Dinis, Luis
            and Petrov, Dmitri and Parrondo, Juan MR and Rica, Ra{\'u}l A},
  journal = {Nature physics},
  volume = {12},
  number = {1},
  pages = {67--70},
  year = {2016},
  publisher = {Nature Publishing Group UK London},
}

@article{li2024realization,
  title = {Realization of an all-optical underdamped stochastic Stirling engine},
  author = {Li, Chuang and Zhu, Shaochong and He, Peitong and Wang, Yingying and
            Zheng, Yi and Zhang, Kexin and Gao, Xiaowen and Dong, Ying and Hu,
            Huizhu},
  journal = {Physical Review A},
  volume = {109},
  number = {2},
  pages = {L021502},
  year = {2024},
  publisher = {APS},
  url = {https://doi.org/10.1103/PhysRevA.109.L021502},
}

@article{chiang2017electrical,
  title = {Electrical autonomous Brownian gyrator},
  author = {Chiang, K-H and Lee, C-L and Lai, P-Y and Chen, Y-F},
  journal = {Physical Review E},
  volume = {96},
  number = {3},
  pages = {032123},
  year = {2017},
  publisher = {APS},
  url = {https://doi.org/10.1103/PhysRevE.96.032123},
}

@article{dechant2015AllOptical,
  title = {All-{{Optical Nanomechanical Heat Engine}}},
  author = {Dechant, Andreas and Kiesel, Nikolai and Lutz, Eric},
  year = 2015,
  month = may,
  journal = {Phys. Rev. Lett.},
  volume = {114},
  number = {18},
  pages = {183602},
  doi = {10.1103/PhysRevLett.114.183602},
  urldate = {2015-06-09},
}

@article{delic2020Cooling,
  title = {Cooling of a Levitated Nanoparticle to the Motional Quantum Ground
           State},
  author = {Deli{\'c}, Uro{\v s} and Reisenbauer, Manuel and Dare, Kahan and
            Grass, David and Vuleti{\'c}, Vladan and Kiesel, Nikolai and
            Aspelmeyer, Markus},
  year = 2020,
  month = feb,
  journal = {Science},
  volume = {367},
  number = {6480},
  pages = {892--895},
  publisher = {American Association for the Advancement of Science},
  doi = {10.1126/science.aba3993},
  urldate = {2022-06-08},
}

@article{gieseler2018Levitated,
  title = {Levitated {{Nanoparticles}} for {{Microscopic Thermodynamics}}---{{A
           Review}}},
  author = {Gieseler, Jan and Millen, James},
  year = 2018,
  month = apr,
  journal = {Entropy},
  volume = {20},
  number = {5},
  pages = {326},
  doi = {10.3390/e20050326},
  urldate = {2018-05-09},
  copyright = {http://creativecommons.org/licenses/by/3.0/},
  langid = {english},
}

@article{gonzalez-ballestero2021Levitodynamics,
  title = {Levitodynamics: {{Levitation}} and Control of Microscopic Objects in
           Vacuum},
  shorttitle = {Levitodynamics},
  author = {{Gonzalez-Ballestero}, C. and Aspelmeyer, M. and Novotny, L. and
            Quidant, R. and {Romero-Isart}, O.},
  year = 2021,
  month = oct,
  journal = {Science},
  volume = {374},
  number = {6564},
  pages = {eabg3027},
  publisher = {American Association for the Advancement of Science},
  doi = {10.1126/science.abg3027},
  urldate = {2021-10-13},
}

@article{magrini2021Realtime,
  title = {Real-Time Optimal Quantum Control of Mechanical Motion at Room
           Temperature},
  author = {Magrini, Lorenzo and Rosenzweig, Philipp and Bach, Constanze and {
            Deutschmann-Olek}, Andreas and Hofer, Sebastian G. and Hong, Sungkun
            and Kiesel, Nikolai and Kugi, Andreas and Aspelmeyer, Markus},
  year = 2021,
  month = jul,
  journal = {Nature},
  volume = {595},
  number = {7867},
  pages = {373--377},
  issn = {1476-4687},
  doi = {10.1038/s41586-021-03602-3},
  urldate = {2021-07-19},
  copyright = {2021 The Author(s), under exclusive licence to Springer Nature
               Limited},
  langid = {english},
}

@misc{message2025Extremetemperature,
  title = {Extreme-Temperature Single-Particle Heat Engine},
  author = {Message, Molly and Cerisola, Federico and Pritchett, Jonathan D. and
            O'Flynn, Katie and Ren, Yugang and Rashid, Muddassar and Anders,
            Janet and Millen, James},
  year = 2025,
  month = jan,
  number = {arXiv:2501.03677},
  eprint = {2501.03677},
  primaryclass = {cond-mat},
  publisher = {arXiv},
  doi = {10.48550/arXiv.2501.03677},
  urldate = {2025-01-08},
  archiveprefix = {arXiv},
  langid = {english},
}

@article{rossnagel2016Singleatom,
  title = {A Single-Atom Heat Engine},
  author = {Ro{\ss}nagel, Johannes and Dawkins, Samuel T. and Tolazzi, Karl N.
            and Abah, Obinna and Lutz, Eric and {Schmidt-Kaler}, Ferdinand and
            Singer, Kilian},
  year = 2016,
  month = apr,
  journal = {Science},
  volume = {352},
  number = {6283},
  pages = {325--329},
  issn = {0036-8075, 1095-9203},
  doi = {10.1126/science.aad6320},
  urldate = {2016-04-17},
  copyright = {Copyright \copyright{} 2016, American Association for the
               Advancement of Science},
  langid = {english},
  pmid = {27081067},
}

@article{tebbenjohanns2021Quantum,
  title = {Quantum Control of a Nanoparticle Optically Levitated in Cryogenic
           Free Space},
  author = {Tebbenjohanns, Felix and Mattana, M. Luisa and Rossi, Massimiliano
            and Frimmer, Martin and Novotny, Lukas},
  year = 2021,
  month = jul,
  journal = {Nature},
  volume = {595},
  number = {7867},
  pages = {378--382},
  publisher = {Nature Publishing Group},
  issn = {1476-4687},
  doi = {10.1038/s41586-021-03617-w},
  urldate = {2022-08-21},
  copyright = {2021 The Author(s), under exclusive licence to Springer Nature
               Limited},
  langid = {english},
}

@article{kamba2023Revealing,
  title = {Revealing the {{Velocity Uncertainties}} of a {{Levitated Particle}}
           in the {{Quantum Ground State}}},
  author = {Kamba, M. and Aikawa, K.},
  year = 2023,
  month = oct,
  journal = {Phys. Rev. Lett.},
  volume = {131},
  number = {18},
  pages = {183602},
  publisher = {American Physical Society},
  doi = {10.1103/PhysRevLett.131.183602},
  urldate = {2023-10-31},
}

@article{raynal2023Shortcuts,
  title = {Shortcuts to {{Equilibrium}} with a {{Levitated Particle}} in the {{
           Underdamped Regime}}},
  author = {Raynal, Damien and {de Guillebon}, Timoth{\'e}e and {Gu{\'e}
            ry-Odelin}, David and Trizac, Emmanuel and Lauret, Jean-S{\'e}bastien
            and Rondin, Lo{\"i}c},
  year = 2023,
  month = aug,
  journal = {Phys. Rev. Lett.},
  volume = {131},
  number = {8},
  pages = {087101},
  doi = {10.1103/PhysRevLett.131.087101},
  urldate = {2023-09-24},
}

@article{rondin2017Direct,
  title = {Direct Measurement of {{Kramers}} Turnover with a Levitated
           Nanoparticle},
  author = {Rondin, Lo{\"i}c and Gieseler, Jan and Ricci, Francesco and Quidant,
            Romain and Dellago, Christoph and Novotny, Lukas},
  year = 2017,
  month = dec,
  journal = {Nature Nanotechnology},
  volume = {12},
  number = {12},
  pages = {1130--1133},
  issn = {1748-3395},
  doi = {10.1038/nnano.2017.198},
  urldate = {2018-02-27},
  copyright = {2017 Nature Publishing Group},
  langid = {english},
}

@article{martinez2013Effective,
  title = {Effective Heating to Several Thousand Kelvins of an Optically Trapped
           Sphere in a Liquid},
  author = {Mart{\'i}nez, Ignacio A. and Rold{\'a}n, {\'E}dgar and Parrondo,
            Juan M. R. and Petrov, Dmitri},
  year = 2013,
  month = mar,
  journal = {Phys. Rev. E},
  volume = {87},
  number = {3},
  pages = {032159},
  issn = {1539-3755, 1550-2376},
  doi = {10.1103/PhysRevE.87.032159},
  urldate = {2024-06-03},
  copyright = {http://link.aps.org/licenses/aps-default-license},
  langid = {english},
}

@book{seifert2025Stochastic,
  title = {Stochastic {{Thermodynamics}}},
  author = {Seifert, Udo},
  year = 2025,
  publisher = {Cambridge University Press},
  address = {Cambridge},
  doi = {10.1017/9781009024358},
  urldate = {2026-01-05},
  isbn = {978-1-316-51955-4},
}

@article{arold2018Heat,
  title = {Heat Leakage in Overdamped Harmonic Systems},
  author = {Arold, Dominic and Dechant, Andreas and Lutz, Eric},
  year = 2018,
  month = feb,
  journal = {Physical Review E},
  volume = {97},
  number = {2},
  pages = {022131},
  issn = {2470-0045, 2470-0053},
  doi = {10.1103/PhysRevE.97.022131},
  urldate = {2025-01-09},
  langid = {english},
}

@article{baldassarri2020Engineered,
  title = {Engineered Swift Equilibration of a {{Brownian}} Gyrator},
  author = {Baldassarri, A. and Puglisi, A. and Sesta, L.},
  year = 2020,
  month = sep,
  journal = {Physical Review E},
  volume = {102},
  number = {3},
  pages = {030105},
  publisher = {American Physical Society},
  doi = {10.1103/PhysRevE.102.030105},
  urldate = {2022-07-12},
}

@article{patron2024Minimuma,
  title = {Minimum Time Connection between Non-Equilibrium Steady States: The {{
           Brownian}} Gyrator},
  shorttitle = {Minimum Time Connection between Non-Equilibrium Steady States},
  author = {Patr{\'o}n, A and Plata, C A and Prados, A},
  year = 2024,
  month = nov,
  journal = {Journal of Physics A: Mathematical and Theoretical},
  volume = {57},
  number = {49},
  pages = {495004},
  publisher = {IOP Publishing},
  issn = {1751-8121},
  doi = {10.1088/1751-8121/ad909a},
  urldate = {2026-01-07},
  langid = {english},
}

\end{document}


\title{Supplementary Information for \\ \textit{Role of Inertia on the Performance of Brownian Gyrators}}

\author{Thalyta T. Martins\orcidlink{https://orcid.org/0000-0003-2113-5468}}%
\author{Ines Ben-Yedder}%
\author{Alex Fontana \orcidlink{https://orcid.org/0000-0001-5522-3584}}%
\author{Loïc Rondin \orcidlink{https://orcid.org/0000-0002-4833-2886}}%
\affiliation{%
 Laboratoire Lumière, Matière et Interfaces, Université Paris-Saclay, Gif-sur-Yvette, France
}%

\maketitle

\section{Experimental setup}
\begin{figure}[ht]
    \centering
    \includegraphics[width=0.8\textwidth]{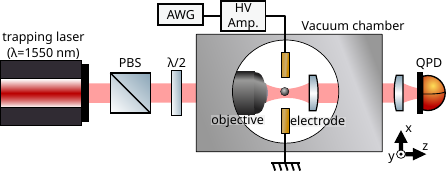}
    \caption{Scheme of the experimental setup. A 106~nm radius nanoparticle is trapped by a $1550 \ \mathrm{nm}$ infrared laser focused using a high-numerical-aperture objective. The light scattered by the particle is collected by a collection lens and detected with a quadrant photodiode (QPD). A polarizer (PBS) and a half-wave plate ($\lambda/2$) allow for the control of the polarization of the laser, and thus the orientation of the anisotropic potential. Anisotropic heating along the $x$-direction is induced by electrical heating generated by an arbitrary waveform generator (AWG), amplified by a high-voltage (HV) amplifier, and applied through the electrodes.}
    \label{fig:setup}
\end{figure}
A schematic of the experimental setup is shown in Figure~\ref{fig:setup}. The optical tweezers system consists of a $\lambda=1550$~nm collimated infrared laser (Keopsys) focused into a vacuum chamber using an $100 \times$, $\mathrm{NA}=0.85$ objective (Olympus, LCPLN100XIR). For trapping, $106 \ \mathrm{nm}$ radius silica nanoparticles (Microparticles GmbH, SiO2-R-L4024-1) diluted in ethanol are introduced into the chamber by nebulizing the solution close to the laser focus. 
Once a particle is stably trapped, its motion is detected using forward-scattering interferometry. The interference signal between the light scattered by the particle and the incoming transmitted beam is collected by a collection lens and directed onto a home-made quadrant photodiode (QPD) to measure the 3D dynamics of the particle. 

The anisotropy of the trapping potential is related to the polarization direction of the laser beam. Thus, to control the orientation of the trapping potential, a Glan–Thompson polarizer (PBS) and a half-wave plate ($\lambda/2$) are placed in the optical path before the vacuum chamber.  A rotation of the half-wave plate by an angle $\alpha$ results in a rotation of the trapping potential by an angle $\phi=2 \alpha$ in the $(x,y)$ plane.

To control the particle's effective temperature, electrical heating is applied. For that, a pair of gold electrodes is positioned near the trapping region along the $x$ direction, with a separation of approximately $1 \ \mathrm{mm}$. A white-noise voltage signal, generated by an arbitrary wave generator (AWG, Keysight 33500B) with tunable amplitude between $0$ and $10 \ \mathrm{V}$ and a bandwidth of $500 \ \mathrm{kHz}$ is amplified by a factor of $50$ using a linear amplifier (Tegam 2340) and then applied to the electrodes.
Data acquisition is performed using a Gage Octave Express CompuScope acquisition card.

\section{Analysis and calibrations of the particle's dynamics}
\begin{figure}[ht]
    \centering
    \includegraphics[width=0.9\textwidth]{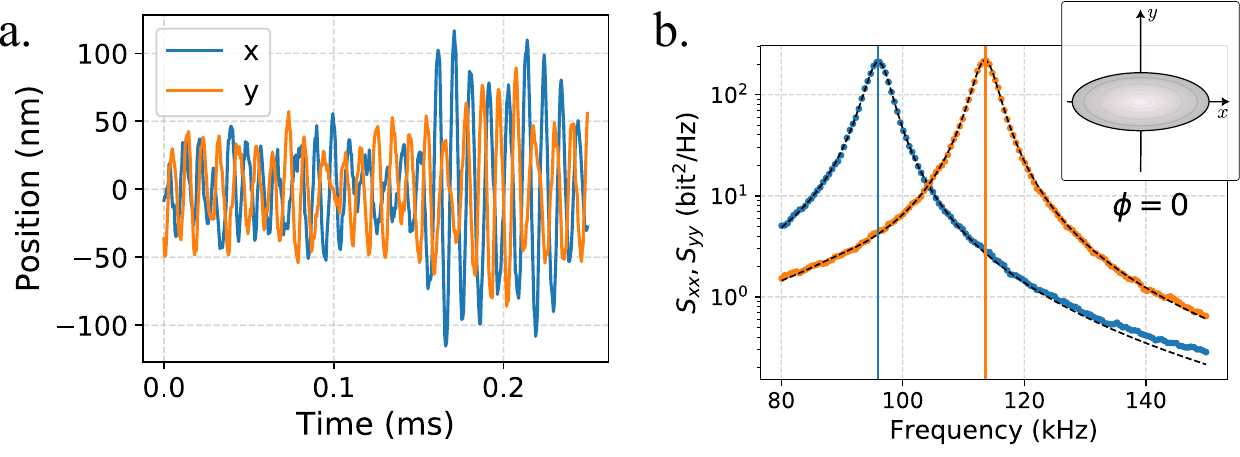}
    \caption{a. Typical trajectory $x(t)$ and $y(t)$ of the trapped nanoparticle. b. corresponding PSDs, $S_{xx}(f)$ and $S_{yy}(f)$. The data are acquired at a gas pressure $p_{\mathrm{gas}}=8 \ \mathrm{mbar}$ (underdamped regime), with the trapping potential aligned with the measurement $x$ axis, i.e. $\phi=0$ (inset). No voltage is applied to the electrodes, corresponding to thermal equilibrium at room temperature, $T=295 \ \mathrm{K}$. Dashed lines correspond to fits using the equation~\eqref{eq:PSD}. The solid vertical lines indicate the resonance frequencies $f_{0,i}$, with $i\in\{x,y\}$. The $x$ and $y$ dynamics are shown in blue and orange, respectively.}
    \label{fig:trajectory}
\end{figure}

All recorded data consist of three independent position time series acquired at a sampling rate of $f_s = 2\ \mathrm{MHz}$ over a total duration of $t_{\mathrm{tot}} = 2.5\ \mathrm{s}$. Each time series represents the measured particle dynamics
\begin{equation}
\bm{r}(t_i) =
\begin{pmatrix}
x(t_i)\\
y(t_i)
\end{pmatrix}, \quad \text{with} \ t_i = i/f_s \ \text{and} \ i=1,\dots,f_s t_{\mathrm{tot}}.
\end{equation}
Figure~\ref{fig:trajectory}a shows a typical trajectory along $x$ and $y$, obtained at a pressure of $p_\text{gas} = 8\ \mathrm{mbar}$ and room temperature (no voltage applied to the electrodes), with $\phi = 0$, \textit{i.e.} with the major axis of the potential aligned along the  $x$-axis (inset of Figure~\ref{fig:trajectory}b).

To evaluate the gyrator angular momentum, $\mathcal{\bm{L}}$ as well as the particle rotation probability current, $\bm{j}(x,y)$, it is necessary to also compute the particle velocity. The velocity $\bm{v}$ is obtained from the discrete time difference of the position trajectories $\bm{r}$. 

For calibration purposes, we also compute the power spectral density (PSD) of the particle trajectories. The PSDs are computed using Welch's method~\cite{welch1967use}. The fast Fourier transform is performed with $N_{\rm FFT} = 2^{12}$ points, and a segment overlap of $0.9 \ N_{\rm FFT}$ is used. In all the analyses, PSDs are calculated for $x$ and $y$ dynamics, and separately for each of the three time series, and the final PSD corresponds to the average of these spectra, as illustrated in Figure~\ref{fig:trajectory}b.

Assuming a harmonic trap at frequency $f_{0,i}=\Omega_{0,i}/(2\pi)$, and equilibrium, the power spectral density (PSD) of the levitated particle's position is given by
\begin{equation}
S_{ii}(f) = \frac{A'}{\pi},
\frac{2 f_{0,i}^2 \Gamma}{\left(f^2 - f_{0,i}^2\right)^2 + f^2 \Gamma^2},
\label{eq:PSD}
\end{equation}
where  $i\in\{x,y\}$,  $\Gamma$ is the system damping rate, $A' = C_i^2 k_B T /(m \Omega_{0, i})$, with $C_i$ the calibration factor converting the measured voltage signal into displacement units (meters), $k_B$ is the Boltzmann constant, $T$ is the bath temperature, and $m$ is the particle mass.

\subsection{Calibration of resonant frequencies and damping}

To calibrate the resonance frequencies $f_{0,i}$ and the damping rate $\Gamma$ of the system, particle trajectories are acquired with the trapping potential aligned at $\phi=0$, i.e., with the major axis of the potential oriented along the laboratory $x$ direction (inset of Fig.~\ref{fig:trajectory}b), and with no voltage applied to the electrodes. Under these conditions, the system is at thermal equilibrium with the surrounding gas at room temperature.
\begin{figure}[ht]
    \centering
    \includegraphics[width=0.5\textwidth]{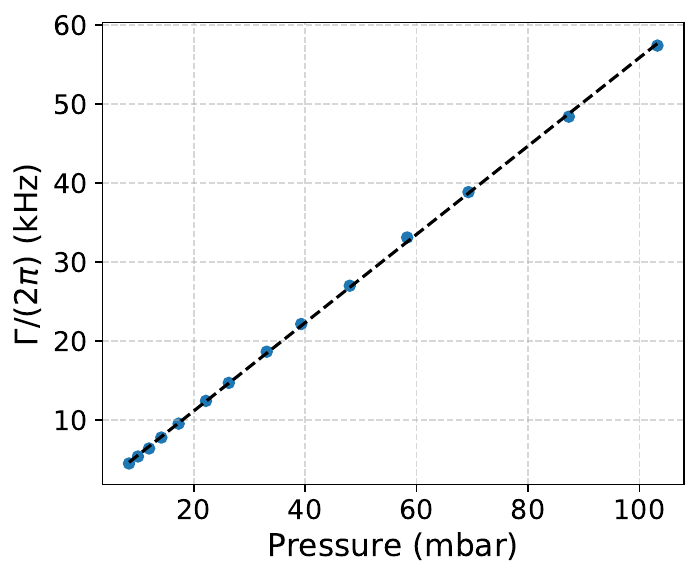}
    \caption{System damping as a function of the gas pressure in the vacuum chamber. The black dashed line is a linear fit.}
    \label{fig:damping_vs_pressure}
\end{figure}

The PSD is computed and fitted using equation~\eqref{eq:PSD}. This procedure is repeated for all investigated vacuum pressures $p_\text{gas}$. From these measurements, we estimate the resonance frequencies to be $f_{0,x}=96.4 \pm 0.4\ \mathrm{kHz}$ and $f_{0,y}=114.2 \pm 0.5\ \mathrm{kHz}$\footnote{The reported resonance frequencies corresponds to the average result obtained at different pressures, and the associated uncertainty is given by the standard deviation of the measurements divided by $\sqrt{N}$, where $N$ is the number of measurements.}. This is equivalent to an average stiffness of $k = 4.81 \ \mathrm{pN/\mu m}$ and a trap asymmetry $u  = 0.8 \ \mathrm{pN/\mu m}$ as mentioned in the main text. 
The linear dependence of the damping as a function of the gas pressure is highlighted in Fig.~\ref{fig:damping_vs_pressure}.


\subsection{Estimating the calibration factor}

From the same data set ($\phi=0$) presented above, one can, in principle, extract the calibration factors $C_x$ and $C_y$. However, we observed a small variation in these calibration factors with the orientation of the $\lambda/2$ plate, which we attribute to small shifts in the trap's equilibrium position. 

\begin{figure}[ht]
    \centering
    \includegraphics[width=0.5\textwidth]{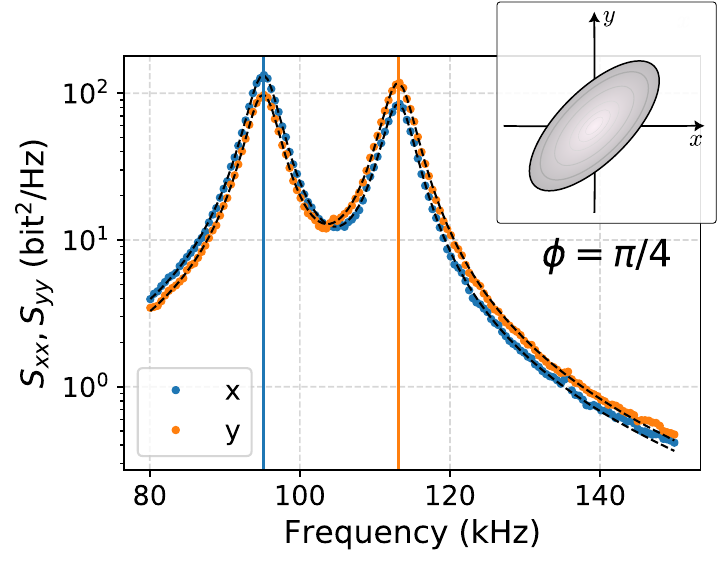}
    \caption{Typical PSDs, $S_{xx}(f)$ and $S_{yy}(f)$, of a trapped particle with $\phi = \pi/4$ (inset). The data are acquired at a gas pressure $p_{\mathrm{gas}}=8 \ \mathrm{mbar}$ (underdamped regime). The electrodes are switched off, corresponding to thermal equilibrium at room temperature, $T=295 \ \mathrm{K}$. The solid vertical lines indicate the resonance frequencies $f_{0,1}$ and $f_{0,2}$ correspond to the first and second peaks, respectively. The $x$ and $y$ dynamics are shown in blue and orange, respectively.}
    \label{fig:psd22}
\end{figure}

Thus, we compute the calibration factor from the data set corresponding to the gyrator configuration, namely $\phi = \pi/4$ for the data presented in the main text.   It is important to note that these measurements are still performed with no voltage applied to the electrodes (in equilibrium at room temperature). Since the system is stable over the pressure range, we use the data set acquired at a low pressure ($p_\text{gas} = 8$~mbar). 

Under these conditions, two well-separated peaks are visible in the PSD, as shown in Fig.~\ref{fig:psd22}. This behavior arises from the projection of the two principal axes of the trapping potential ($x_\text{pot},y_\text{pot}$) onto the fixed axis of the laboratory frame ($x,y$). 

To account for this effect, the PSD is fitted with a sum of two PSD model functions (Eq.~\ref{eq:PSD}). The calibration factors are then extracted as
\begin{equation}
C_i = \sqrt{\frac{\left(A_1' f_{0,1}^2 + A_2' f_{0,2}^2\right)m}{k_B T}}.
\end{equation}
Here, $A_1'$ and $f_{0,1}$ denote the amplitude and resonance frequency associated with the first peak, while $A_2'$ and $f_{0,2}$ correspond to the second peak. 

The resulting calibration factors, $C_x = 41  \ \mathrm{bit/nm}$ and $C_y = 42  \ \mathrm{bit/nm}$ are used throughout the analysis.

\subsection{Calibration of the hot temperature}

\begin{figure}[ht]
    \centering
    \includegraphics[width=0.9\textwidth]{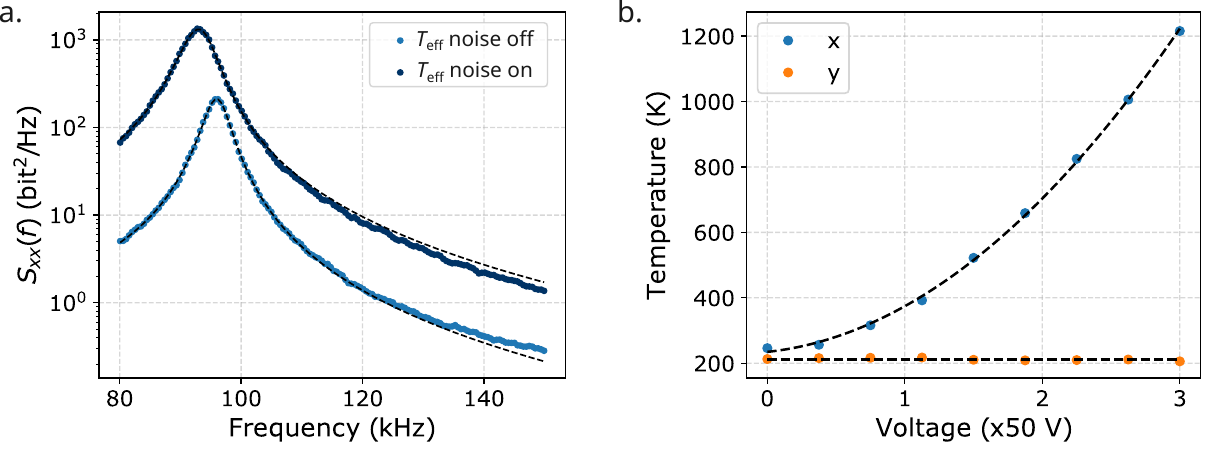}
    \caption{a. Typical PSDs, $S_{xx}(f)$, measured without (blue) and with (darker blue) voltage applied to the electrodes. b. Effective center-of-mass temperatures along the $x$ (blue) and $y$ (orange) directions as a function of the voltage amplitude applied to the electrodes. While the temperature along $x$ exhibits a quadratic dependence, the temperature along $y$ remains constant. Black dashed lines indicate fits to the data.}
    \label{fig:psd_with_noise} 
\end{figure}

To generate an anisotropic bath for the particle, a controlled Gaussian white-noise electric signal is applied to the electrodes, as mentioned previously. The applied noise causes a high effective temperature along the  $x$ direction that results in an increase of the area under the PSD curve (Fig.~\ref{fig:psd_with_noise}.a ). From the PSD evolution, we can directly extract the temperature by assuming it is at room temperature ($295 \ \mathrm{K}$) when no voltage is applied to the electrodes. It is important to mention that the calibration of the effective center-of-mass temperature is performed with the trapping potential aligned at $\phi=0$, since for $\phi\neq0$ the gyrator is active and the resulting dynamics would not allow for a reliable temperature calibration. 

The effective temperature has a quadratic dependence on the voltage amplitude, $V$, as shown in Fig. \ref{fig:psd_with_noise}.b. It follows the equation \cite{militaru2021kovacs}: 
\begin{equation}
    T_{\mathrm{eff}} = \frac{1}{2m\Gamma k_B b} \left(\frac{qV}{d}\right)^2
\end{equation}
being $q$ the charge of particle, $d \approx 1 \ \mathrm{mm}$ the distance between the electrodes, and $b=500 \ \mathrm{kHz}$ the bandwidth of the signal.

Therefore, to maintain a fixed effective temperature across different damping rates, the calibration procedure illustrated in Fig.~\ref{fig:psd_with_noise}b is performed independently for each vacuum chamber pressure. From a quadratic fit, we determine the voltage required to reach a temperature eight times above the room temperature for the results presented in this work. For all the pressures, we have an estimated temperature of
$T_{\text{eff}} = 2360 \pm 40 \ \mathrm{K}$ in the $x$ direction~\footnote{Note that this temperature corresponds to the average result obtained at different pressures, and the associated uncertainty is given by the standard deviation of the measurements divided by $\sqrt{N}$, where $N$ is the number of measurements.} while the $y$ direction remains at $295 \ \mathrm{K}$ (no temperature increase was detected in this direction). 

\newpage
\section{Effect of $\phi$ on the Brownian Gyrator}

\begin{figure}[ht]
    \centering
    \includegraphics[width=0.9\textwidth]{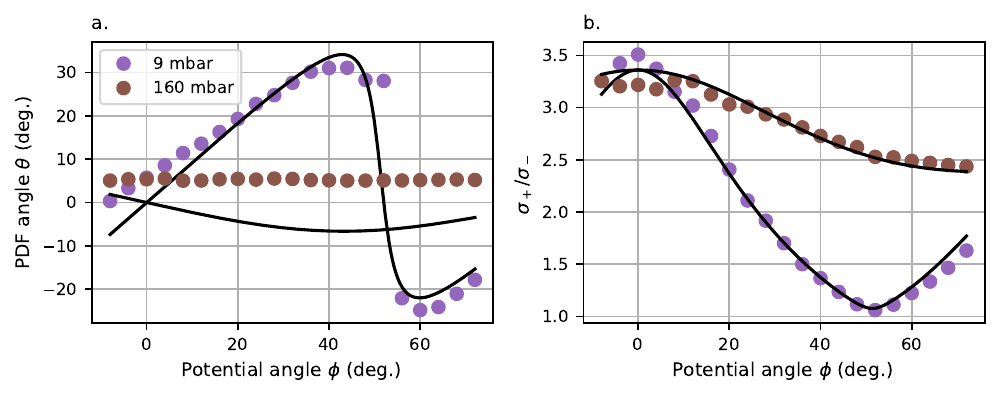}
    \caption{a. Measured angle $\theta$ and b. aspect ratio of the standard deviations $\sigma_+/\sigma_-$ of the principal axis of the PDF as a function of the potential orientation angle $\phi$ (controlled by rotating the $\lambda/2$ plate). Results for low ($9 \ \mathrm{mbar}$) and high ($160 \ \mathrm{mbar}$) pressures are presented in purple and brown, respectively. Here, the results were obtained, as discussed in the main text, for  $T_\mathrm{hot}=8 \ T_\mathrm{cold}$. The solid black line represents the theoretical model. }
    \label{fig:angle_misalignment}
\end{figure}

In our experiment, the reference frame is defined by the electrode orientation, which determines the direction along which the particle's effective temperature increases. The electrodes' direction is fixed and cannot be modified. 
To determine the position where $x_{\mathrm{pot}}$ aligns with the hot (electrodes) axis, i.e., $\phi=0$, we analyzed the Brownian gyrator response (with voltage applied to the electrodes) as a function of the half-wave plate rotation. Figure~\ref{fig:angle_misalignment} shows the measured rotation angle $\theta$ (a) and the aspect ratio of the standard deviations (b) along the principal axes of the PDF, for both low  ($ 9 \ \mathrm{mbar}$) and high ($ 160 \ \mathrm{mbar}$) pressure conditions and $T_\mathrm{hot}=8 \ T_\mathrm{cold}$ for both measurements. When the angle $\theta$ is precisely the same at both pressures, we can assume the gyrator has no effect, i.e., $\phi = 0$ truly.
This reference orientation of the half-wave plate defines $\phi=0$, while further rotation of the plate by $45^\circ/2$ corresponds to $\phi=45^\circ$. 

It is worth noting that a small misalignment between the direction of the electrodes and the laboratory reference frame is experimentally observed. To account for that, all measured trajectories and angles are rotated by a constant correction angle of $-4^\circ$.

\bibliographystyle{plain}  
\bibliography{ref} 